\documentclass[10pt,journal]{IEEEtran}

\author{\authorblockN{Mustafa Cenk Gursoy and Qingyun Wang}
\thanks{M. C. Gursoy is with the Department of Electrical
Engineering, University of Nebraska-Lincoln, Lincoln, NE 68588. Q. Wang was also with the the Department of Electrical
Engineering, University of Nebraska-Lincoln. She is now with Automated Precision Inc., Rockville, MD 20850.
(e-mails:gursoy@engr.unl.edu, wqyjiyang@hotmail.com)}
\thanks{This work was supported in part by the NSF CAREER
Grant CCF-0546384.}}

\usepackage{amsmath}
\usepackage{amssymb}
\usepackage[dvips]{graphicx}
\usepackage{setspace}
\usepackage{epsfig}

\newcommand{\h}{\mathbf{h}}

\newcommand{\R}{{\mathbf{R}}}

\newcommand{\tth}{{^\text{th}}}

\newcommand{\figsize}{0.48}

\newtheorem{lemma1}{Lemma}

\begin{document}

\title{Diversity Analysis of Peaky FSK Signaling in Fading Channels}

\date{}

\maketitle

\begin{abstract}
Error performance of noncoherent detection of on-off frequency shift
keying (OOFSK) modulation over fading channels is analyzed when the
receiver is equipped with multiple antennas. The analysis is
conducted for two cases: 1) the case in which the receiver has the channel distribution knowledge only; and 2) the case in which the receiver perfectly knows the fading magnitudes. For
both cases, the maximum a posteriori probability (MAP) detection
rule is derived and analytical probability of error expressions are
obtained. Numerical and simulation results indicate that for sufficiently low duty
cycle values, lower error probabilities with respect to FSK
signaling are achieved. Equivalently, when compared to FSK
modulation, OOFSK with low duty cycle requires less energy to
achieve the same probability of error, which renders this modulation
a more energy efficient transmission technique. Also, through
numerical results, the impact of number of antennas, antenna
correlation, duty cycle values, and unknown channel fading on the performance are
investigated.

\emph{Index Terms:} Diversity, duty factor, fading channels, frequency-shift keying, MAP detection, multiple antennas, on-off keying, probability of error.

\end{abstract}


\section{Introduction}

Frequency-shift keying (FSK) is a modulation format that is
well-known and well-studied in the communications literature
\cite{proakis}. FSK is an attractive transmission scheme due to its
high energy efficiency and suitability for noncoherent
communications. For instance, under unknown channel conditions,
energy detection can be employed to detect the FSK signals. Indeed,
the analysis of FSK modulation dates back to 1960s (see e.g.,
\cite{Pierce} and \cite{lindsey}). Recently, it has been shown in
\cite{Verdu} that unless the channel conditions are perfectly known
at the receiver, signals that have very high peak-to-average power
ratio is required to achieve the capacity in the low-SNR regime.
This has initiated work on peaky signaling. Luo and M\'edard
\cite{LuoMedard} have shown that FSK with small duty cycle can
achieve rates of the order of capacity in ultrawideband systems with
limits on bandwidth and peak power. In \cite{desmond}, the authors
have studied the error performance of peaky FSK signaling over
multipath fading channels by obtaining upper and lower bounds on the
error probability. In \cite{gursoyoofsk}, on-off frequency-shift
keying (OOFSK) is defined as FSK overlaid on on-off keying, and its
capacity and energy efficiency is analyzed. Note that OOFSK can be
seen as joint pulse position modulation (PPM) and FSK. In this
signaling, peakedness is introduced in both time and frequency. The
error performance of OOFSK signaling when the transmitter and
receiver are each equipped with a single antenna is recently studied
in \cite{qingyun}.

One of the important techniques to improve the performance in
wireless communications is to use multiple antennas to achieve
diversity gain. Considerable amount of work has been done on
multiple reception channels \cite{marvin}. In \cite{lindsey}, it is
shown for binary and $M$-ary signaling over Rician fading channels
that increasing the number of reception channels can improve the
error performance significantly. By finding the probability
distribution function of the instantaneous SNR in flat fading
multi-reception channels and substituting it into the probability of
error expressions of PAM, PSK and QAM over an AWGN channel, the authors
in \cite{hao} obtained expressions for the average probability of
error of multi-reception fading channels. In \cite{veer}, the
probability of error of BPSK over Rician fading multi-reception
channels is given and extensions to other modulation techniques are
discussed. In \cite{yunfei}, average symbol error rate of selection
diversity of $M$-ary FSK modulated signal transmitted over fading
channels is studied.

In this paper, the error performance of noncoherent detection of OOFSK over multiple
reception Rician fading channels is studied. In Section
\ref{sec:model}, the system model is presented. In Section \ref{sec:noncoherent}, the error performance in unknown Rician fading
channels is studied. In Section \ref{sec:coherent}, we
investigate the error performance in known fading
channels. Finally, Section \ref{sec:conclusion} includes our
conclusions.

\section{System Model}\label{sec:model}

OOFSK modulation is employed at the transmitter. In OOFSK modulation, the transmitted signal
during the symbol interval $0 \le t \le T_s$ can be expressed as
\begin{align}
s_m(t)=\left\{\begin{array}{ll}\sqrt{\frac{P}{v}}e^{j(w_mt+\theta_m)}
& m=1,2,3,\ldots M\\0  & m=0
\end{array}\right.
\end{align}
where $w_m$ and $\theta_m$ are the frequency in radians per second
and phase, respectively, of the signal $s_m(t)$ when $m \neq 0$.
Note that we have $M$ FSK signals and a zero signal denoted by
$s_0(t)$. The frequencies of the FSK signals are chosen so that the
signals are orthogonal. Since noncoherent detection of OOFSK signals is considered throughout the paper, the phases $\theta_m$ can be arbitrary. It is assumed that an FSK signal $s_m(t)$,
$m \neq 0$, is transmitted with a probability of $\frac{v}{M}$ while
$s_0(t)$ is transmitted with a probability of $1-v$ where $v$ is the
duty cycle of the transmission. With these definitions, it is easily
seen that $P$ and $\frac{P}{v}$ are the average and peak powers,
respectively, of the modulation technique.  Therefore, constraints on the peak power will impose lower bounds on the values that the duty cycle parameter $v$ can assume.

\par The receiver is equipped with $L$ antennas that enable
the multiple reception of the transmitted signal. If, without loss
of generality, we assume that $s_k(t)$ is the transmitted signal,
the received signal at the $l^{\text{th}}$ antenna is
\begin{equation} \label{eq:receivedsignal}
r_l(t)=h_ls_k(t)+n_l(t)\quad l=1,2,\ldots,L
\end{equation}
where $h_l$ is the fading coefficient of the  $l^{\text{th}}$
reception channel and $n_l(t)$ is a white Gaussian noise process
with single-sided spectral density of $N_0$. It is assumed that the
additive Gaussian noise components at different antennas are
independent. Furthermore, the received signal model
(\ref{eq:receivedsignal}) presumes that the fading is frequency-flat
and slow enough so that the fading coefficients stay constant over
one symbol duration.

Following each antenna, there is a bank of $M$ correlators, each
correlating the received signal with one of the orthogonal
frequencies.
The output of the $m^{\text{th}}$ correlator employed after the
$l^{\text{th}}$ antenna is given by
\begin{align}
Y_{l,m}&=\frac{1}{\sqrt{N_0T_s}}\int_{0}^{T_s}r_l(t)e^{-jw_mt}dt
\\
&=\left\{\begin{array}{ll}A h_l
e^{j\theta_m}+n_{l,m} &m=k\\
n_{l,m} &m\neq k \end{array}\right.\label{eq:output}
\end{align}
where $n_{l,m}$ is a circularly symmetric complex Gaussian random
variable with zero-mean and a variance of $1$ (due to normalization with $\frac{1}{\sqrt{N_0T_s}}$) and for notational
convenience, we have defined $A=\sqrt{\frac{PT_s}{vN_0}}$. Since the
frequencies are orthogonal and the additive Gaussian noise is
independent at each antenna, $\{n_{l,m}\}$ for $l \in
\{1,\ldots,L\}$ and $m \in \{1,\ldots,M\}$ forms an independent and
identically distributed (i.i.d.) sequence. Note also that
$R_{l,m}=|Y_{l,m}|^2$ gives the energy present in the
$m^{\text{th}}$ frequency at the $l^{th}$ antenna.

\section{Noncoherent Detection of OOFSK with Channel Distribution Knowledge Only }
\label{sec:noncoherent}

\subsection{Detection Rule}

In this section, we assume that the realizations of
the fading coefficients $\{h_l\}$ are unknown at both the receiver
and transmitter. The receiver is only equipped with the knowledge of
the statistics of $\{h_l\}$. We further assume that $\{h_l\}$ are
i.i.d. complex Gaussian random variables with $E\{h_l\} = d_l$ and
$Var\{h_l\} = \sigma^2$. Hence, we consider a Rician fading channel model which specializes to Rayleigh fading when $d_l = 0$. With the described channel statistics, conditioned on $s_k(t)$ being
the transmitted signal, $Y_{l,m}$ is a complex Gaussian random
variable with
\begin{align} \label{eq:Ynoncoh}
\begin{split}
E\{Y_{l,m}|s_k\} &= \left\{
\begin{array}{ll}
Ad_le^{j\theta_k} & m=k
\\
0 & m\neq k
\end{array}\right.,
\text{ and}
\\
Var\{Y_{l,m}|s_k\}&= \left\{
\begin{array}{ll}
A^2 \sigma^2 + 1 & m = k
\\
1 & m \neq k
\end{array}
\right..
\end{split}
\end{align}
The receiver is assumed to perform noncoherent energy detection and
therefore compute $R_{l,m} = |Y_{l,m}|^2$ after the correlator.
$R_{l,m}$ is chi-square distributed with the following conditional
probability density function (pdf) given the transmitted signal
$s_k$ \cite{proakis}:
\begin{align} \label{eq:Rlmpdfnoncoh}
f_{R_{l,m}|\,s_k}(R_{l,m})=\left\{\begin{array}{ll}
\!\!\!\!\frac{1}{\sigma_y^2}\,
    e^{-\frac{R_m+A^2|d_l|^2}{\sigma_y^2}}I_{0}\left(\frac{2A|d_l|\sqrt{R_m}}{\sigma_y^2}\right)&\!m=k\\
\!\!\!\!e^{-R_{l,m}}&\!m\neq k\end{array}\right.
\end{align}
where $\sigma_y^2 = A^2 \sigma^2 + 1$ and $I_0(\cdot)$ is the zeroth order modified Bessel function of the first kind. It is assumed that the receiver, using equal gain combining,
combines the energies of the $m^{\text{th}}$ frequency component at
each antenna, i.e., computes the total energy
$$R_m=\sum_{l=1}^LR_{l,m}.$$
Since the additive noise components and fading coefficients at different antennas are independent, $R_m$ is a sum of
independent chi-square random variables, and is itself also
chi-square distributed with $2L$ degrees of freedom. Hence, the
conditional pdf of $R_m$ is given by \cite{proakis}
\begin{align}\label{pdf2}
f_{R_m|s_k}(R_m)\!=\!\left\{\begin{array}{ll}
\!\!\!\!\frac{1}{\sigma_y^2}\left(\frac{R_m}{\xi}\right)^{\frac{L-1}{2}}
e^{-\frac{R_m+\xi}{\sigma_y^2}}I_{L-1}\left(\frac{2\sqrt{R_m\xi}}{\sigma_y^2}\right)
&\!\!m=k\\ \!\!\!\!\frac{R_m^{L-1}}{\Gamma(L)}\,e^{-R_m}&\!\!m \neq k
\end{array}\right.
\end{align}
where $\xi=A^2\sum_{l=1}^L |d_l|^2$, $I_{L-1}(\cdot)$ is the $(L-1)^{\text{th}}$ order modified Bessel
function of the first kind, and $\Gamma(\cdot)$ is the gamma
function.

The receiver employs maximum a posteriori probability (MAP)
criterion to detect the transmitted signals.  Let
$\R=[R_1,R_2,\ldots,R_M]$ be the vector of energy values
corresponding to each frequency. Since the noise components
$n_{l,m}$ are independent for different $m \in \{1,\ldots,M\}$, the
components of $\R$ are mutually independent. Hence, the conditional
pdf of $\R$ is given by the product of the marginal pdf's (see equation \eqref{eq:pdf_R_noncoh} on the next page).
\begin{figure*}
\begin{align}
f_{\R|s_k}(\R)=\left\{\begin{array}{ll}\frac{1}{\sigma_y^2}\left(\frac{R_k}{\xi}\right)^{\frac{L-1}{2}}
e^{-\frac{R_k+\xi}{\sigma_y^2}}I_{L-1}\left(\frac{2\sqrt{R_k\xi}}{\sigma_y^2}\right)\prod_{\substack{n=1 \\
n\neq k}
}^M \frac{R_n^{L-1}e^{-R_n}}{\Gamma(L)}, &  k\neq 0\\
\frac{1}{\left[\Gamma(L)\right]^{M}}\prod_{n=1}^MR_n^{L-1}e^{-R_n},
& k=0\end{array}\right.. \label{eq:pdf_R_noncoh}
\end{align}
\hrule
\end{figure*}
Note that given the decision variables $\R=[R_1,R_2,\ldots,R_M]$, MAP detection minimizes the error probabilities \cite{proakis}. The MAP decision rule that detects $s_k$ for $k \neq 0$ is
\begin{align}\label{compare1nc}
f_{\R|s_k}>f_{\R| s_m} \quad \forall m\neq 0, k \quad \text{and}
\quad f_{\R| s_k}>\frac{M(1-v)}{v}f_{\R| s_0}
\end{align}
where we have used the fact that the prior probabilities of the
transmitted signals are $p(s_m) = \frac{v}{M}$ for all $m \neq 0$,
and $p(s_0) = (1-v)$. Substituting (\ref{eq:pdf_R_noncoh}) into
(\ref{compare1nc}), we can simplify the decision rule to
\begin{align}\label{com1}
\begin{split}
g_1(R_k)&>g_1(R_m) \quad \forall m \neq k \quad \text{and}
\\
g_1(R_k)&>\frac{M(1-v)\sigma_y^2e^{\frac{\xi}{\sigma_y^2}}
\xi^{\frac{L-1}{2}}}{v(L-1)!}
\end{split}
\end{align}
where
$g_1(R_k) =
R_k^{-\frac{L-1}{2}}e^{\frac{R_kA^2\sigma^2}{\sigma_y^2}}I_{L-1}\left(\frac{2\sqrt{R_k\xi}}{\sigma_y^2}\right)$
with $\xi>0$.
\newline
The following Lemma enables us to further simplify the detection
rule.
\begin{lemma1} \label{lemma1}
The function $$ g_1(x)=x^{-\frac{L-1}{2}}e^{\frac{x
A^2\sigma^2}{\sigma_y^2}}I_{L-1}\left(\frac{2\sqrt{x\xi}}{\sigma_y^2}\right)$$
for $x
> 0$ and $\xi
> 0
$ is a monotonically increasing function of $x$. Moreover,
$$\lim_{x\to 0} g_1(x) =
\frac{\xi^{\frac{L-1}{2}}}{\sigma_y^{2(L-1)}(L-1)!}.$$
\end{lemma1}

\emph{Proof}: The derivative of the $n\tth$ order modified Bessel
function is
$
\frac{dI_n(x)}{dx} = I_{n+1}(x) + \frac{n}{x}I_n(x)
$ \cite[Section 8.48]{tableofintegrals}.
Hence,
\begin{align}
\frac{d I_{L-1}(\frac{2\sqrt{x\xi}}{\sigma_y^2})}{dx} &=
\frac{1}{\sigma_y^2}\sqrt{\frac{\xi}{x}}I_L\left(\frac{2\sqrt{x\xi}}{\sigma_y^2}\right)
+ \frac{L-1}{2x}I_{L-1}\left(\frac{2\sqrt{x\xi}}{\sigma_y^2}\right)
\\
&>
\frac{L-1}{2x}I_{L-1}\left(\frac{2\sqrt{x\xi}}{\sigma_y^2}\right)
\label{eq:besselderivlower}
\end{align}
where we have used the fact that
$\sqrt{\frac{\xi}{x}}I_L\left(\frac{2\sqrt{x\xi}}{\sigma_y^2}\right)>0$
for $x>0$. Then, the derivative of $g_1(\cdot)$ satisfies \eqref{eq:dg1_1} through \eqref{eq:dg1_3} (see next page)
\begin{figure*}
\begin{align}
\frac{dg_1(x)}{dx}&=-\frac{L-1}{2}x^{-\frac{L+1}{2}}e^{\frac{x
A^2\sigma^2}{\sigma_y^2}}I_{L-1}\left(\frac{2\sqrt{x\xi}}{\sigma_y^2}\right)+x^{-\frac{L-1}{2}}e^{\frac{x
A^2\sigma^2}{\sigma_y^2}}\frac{dI_{L-1}\left(\frac{2\sqrt{x\xi}}{\sigma_y^2}\right)}{dx} 
+\frac{A^2\sigma^2}{\sigma_y^2}
x^{-\frac{L-1}{2}}e^{\frac{x
A^2\sigma^2}{\sigma_y^2}}I_{L-1}\left(\frac{2\sqrt{x\xi}}{\sigma_y^2}\right)\label{eq:dg1_1}
\\
&> -\frac{L-1}{2}x^{-\frac{L+1}{2}}e^{\frac{x
A^2\sigma^2}{\sigma_y^2}}I_{L-1}\left(\frac{2\sqrt{x\xi}}{\sigma_y^2}\right)+x^{-\frac{L-1}{2}}e^{\frac{x
A^2\sigma^2}{\sigma_y^2}}\frac{L-1}{2x}I_{L-1}\left(\frac{2\sqrt{x\xi}}{\sigma_y^2}\right)
+\frac{A^2\sigma^2}{\sigma_y^2}
x^{-\frac{L-1}{2}}e^{\frac{x
A^2\sigma^2}{\sigma_y^2}}I_{L-1}\left(\frac{2\sqrt{x\xi}}{\sigma_y^2}\right)
\label{eq:dg1lower}
\\
&=\frac{A^2\sigma^2}{\sigma_y^2} x^{-\frac{L-1}{2}}e^{\frac{x
A^2\sigma^2}{\sigma_y^2}}I_{L-1}\left(\frac{2\sqrt{x\xi}}{\sigma_y^2}\right)
>0 \quad \text{for } x >0 \label{eq:dg1_3}
\end{align}
\hrule
\end{figure*}
proving that $g_1(x)$ is a monotonically increasing function of
$x>0$. Note that (\ref{eq:dg1lower}) follows from the lower bound
expression provided in (\ref{eq:besselderivlower}). Finally, the
limit expression can be easily shown by using the series expansion
$I_{L-1}\left(\frac{2\sqrt{x\xi}}{\sigma_y^2}\right) =
x^{\frac{L-1}{2}}
\left(\frac{\xi}{\sigma_y^4}\right)^{\frac{L-1}{2}}
\sum_{k=0}^\infty \frac{\left(\frac{\xi x}{\sigma_y^4}\right)^k}{k!
\Gamma(L + k)}$ \cite[Section 8.44]{tableofintegrals}.\hfill $\blacksquare$

With the result of Lemma \ref{lemma1}, the decision rule in
(\ref{compare1nc}) now simplifies to
\begin{align}\label{ncfinalcompare1}
\begin{split}
R_k&>R_m \quad \forall m \neq k \quad \text{and}
\\
R_k&>\tau_1 =
\left\{
\begin{array}{ll}
g_1^{-1}(T_1) & \text{if } T_1 \ge
\frac{\xi^{\frac{L-1}{2}}}{\sigma_y^{2(L-1)}(L-1)!}
\\
0 & \text{otherwise}
\end{array}\right.
\end{split}
\end{align}
where
$T_1=\frac{M(1-v)\sigma_y^2\xi^{\frac{L-1}{2}}e^{\frac{\xi}{\sigma_y^2}}}{v(L-1)!}.$
Since
$g_1(\cdot)$ is a monotonically increasing function, $g_1^{-1}(T_1)$
is well-defined for $T_1 \ge \frac{\xi^{\frac{L-1}{2}}}{\sigma_y^{2(L-1)}(L-1)!}$. Finally, note that $s_0$ is the detected signal if $R_k < \tau_1$ for all
$k$.

\subsection{Probability of Error} \label{subsec:errornoncoh}
In this section, we analyze the error probability of OOFSK
modulation when MAP detection is used at the receiver. Suppose
without loss of generality that $s_1(t)$ is the transmitted signal.
Then the correct detection probability is
\begin{align}
P_{c,1}&=P(R_1>R_2,R_1>R_3,\ldots,R_1>R_M,R_1>\tau_1|s_1)\nonumber\\
&=\int_{\tau{_1}}^\infty\left(\int_0^{x}f_{R_2|
s_1}(t)\,dt\right)^{M-1}f_{R_1| s_1}(x)\,dx \nonumber
\\
&=\int_{\tau{_1}}^\infty\left(\int_0^{x}\frac{t^{L-1}}{\Gamma(L)}e^{-t}dt\right)^{M-1}f_{R_1|
s_1}(x)\,dx\nonumber.
\end{align}
From \cite{proakis}, we know that $
\int_0^x\frac{1}{\Gamma(L)}t^{L-1}e^{-t}dt=1-e^{-x}\sum_{l=0}^{L-1}\frac{x^l}{l!}
$. Therefore, the correct detection probability can now be expressed
as 
\begin{align}
P_{c,1}&=\int_{\tau_1}^\infty\left[1-e^{-x}\sum_{l=0}^{L-1}\frac{x^l}{l!}\right]^{M-1}f_{R_1|
 s_1}(x)dx
 \\
 &= \int_{\tau_1}^\infty
\sum_{n=0}^{M\!-\!1}(-1)^n\left(\!\!\!\begin{array}{cc}M\!-\!1\\n\end{array}\!
\!\!\right)\!\!\left[\sum_{l=0}^{L-1}\!\!\frac{x^l}{l!}e^{-x}\right]^n\!\!\!f_{R_1| s_1}(x)dx \label{eq:Pc1noncoh}
\end{align}
where the rightmost expression is obtained using the binomial
theorem. Moreover, multinomial expansion provides
$
\left[\sum_{l=0}^{L-1}\frac{x^l}{l!}e^{-x}\right]^n=e^{-nx}\sum_{i=0}^{n(L-1)}c_{in}x^i
$ where $c_{in}$ is the coefficient of $x^i$ in the expansion.
$c_{in}$ can be evaluated from the recursive equation
$
c_{in}=\sum_{q=i-L+1}^i\frac{c_{q(n-1)}}{(i-q)!}1_{[0,(n-1)(L-1)]}(q) \text{ where } 1_{[a,b]}(q)= 1 \text{ for } a\leq q\leq b, \text{ and is equal to } 0 \text{ otherwise } \cite{marvin}.
$
Using the multinomial expansion, $P_{c,1}$ can be expressed as in \eqref{eq:errorprobhyper} on the next page.
\begin{figure*}
\begin{align}
P_{c,1}&=\sum_{n=0}^{M-1}(-1)^n\left(\!\!\!\begin{array}{cc}M-1\\n\end{array}\!\!\!\right)
         \sum_{i=0}^{n(L-1)}\!\!c_{in}\int_{\tau_1}^\infty\!\!x^ie^{-nx}f_{R_1|s_1}(x)dx \\
&=\sum_{n=0}^{M-1}(-1)^n\left(\begin{array}{cc}M-1\\n\end{array}\right)
\sum_{i=0}^{n(L-1)}c_{in}\int_{\tau_1}^\infty
x^ie^{-nx}\frac{1}{\sigma_y^2}\left(\frac{x}{\xi}\right)^{\frac{L-1}{2}}e^{-\frac{x+\xi}{\sigma_y^2}}I_{L-1}\left(\frac{2\sqrt{x\xi}}{\sigma_y^2}\right)dx
\\
\hspace{1.5cm}&=\sum_{n=0}^{M-1}(-1)^n\left(\begin{array}{cc}M-1\\n\end{array}\right)\sum_{i=0}^{n(L-1)}c_{in}
\frac{\xi^{-\frac{L-1}{2}}e^{-\frac{\xi}{\sigma_y^2}}}{\sigma_y^2}
\left[\frac{\xi^{\frac{L-1}{2}}(i+L)!}{2(1+n\sigma_y^2)^{\frac{i+L}{2}}
\sigma_y^{L-2-i}L!}F\left(-i,L,\frac{\xi}{\sigma_y^2(1+n\sigma_y^2)}\right)\right.\nonumber\\
&\hspace{5.5cm}\left.\times
e^{\frac{\xi}{\sigma_y^2(1+n\sigma_y^2)}}-\int_0^{\tau_1}
x^{\frac{2i+L-1}{2}}e^{-\frac{1+n\sigma_y^2}{\sigma_y^2}x}I_{L-1}\left(\frac{2\sqrt{x\xi}}{\sigma_y^2}\right)dx\right]. \label{eq:errorprobhyper}
\end{align}
\hrule
\end{figure*}
(\ref{eq:errorprobhyper}) is obtained by noting the following integration result from \cite[Equation (81)]{lindsey}:
\begin{align}
\int_0^\infty &e^{-a^2 x^2} x^{\mu-1}I_\nu(bx) dx \nonumber
\\
&= \frac{b^\nu \Gamma\left( \frac{\mu + \nu}{2}\right)}{2^{\nu + 1} a^{\mu + \nu} \Gamma(\nu + 1)}e^{\frac{b^2}{4a^2}} F\left(\frac{\nu-\mu}{2} + 1, \nu + 1; \frac{-b^2}{4a^2}\right)\nonumber
\end{align}
where $F(a,c;z)$ is the confluent hypergeometric function \cite{andrews}.
Note that while the correct detection probability $P_{c,1}$ can be numerically computed using (\ref{eq:Pc1noncoh}), simplifications in numerical integration can be provided by (\ref{eq:errorprobhyper}) in which the integral has finite limits.

If $s_0(t)$ is the transmitted signal, the probability of correct
detection is
\begin{align} \label{eq:Pc0noncoh}
P_{c,0}&=P\left(R_1<\tau_1,\ldots,R_M<\tau_1 |s_0\right)= \left( 1 -
e^{-\tau_1} \sum_{l=0}^{L-1} \frac{\tau_1^l}{l!}\right)^M.
\end{align}
Finally, the average probability of error is $$
P_e=1-(vP_{c,1}+(1-v)P_{c,0}). $$

\begin{figure}
\begin{center}
\includegraphics[width = \figsize\textwidth]{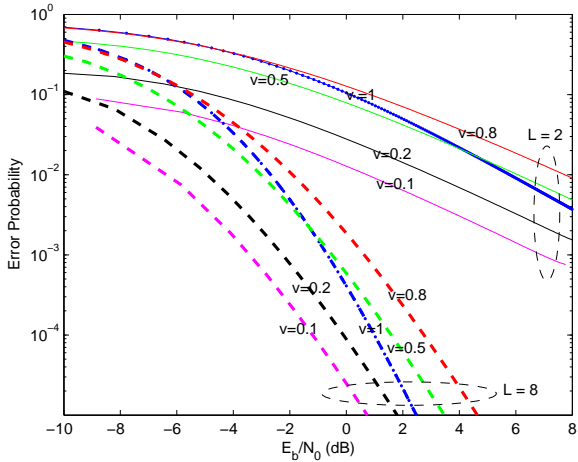}
\caption{Error probability vs. $E_b/N_0$ (dB) for 8-OOFSK signaling over independent Rician fading channels with equal Rician
factor $K=1$. The receiver has only channel distribution knowledge. Solid lines provide the error rates for $L = 2$ channels while dashed lines are for $L = 8$ channels. Duty cycle values are 0.1, 0.2, 0.5, 0.8, 1.}\label{Pe-noncoh-M8}
\end{center}
\end{figure}

\begin{figure}
\begin{center}
\includegraphics[width = \figsize\textwidth]{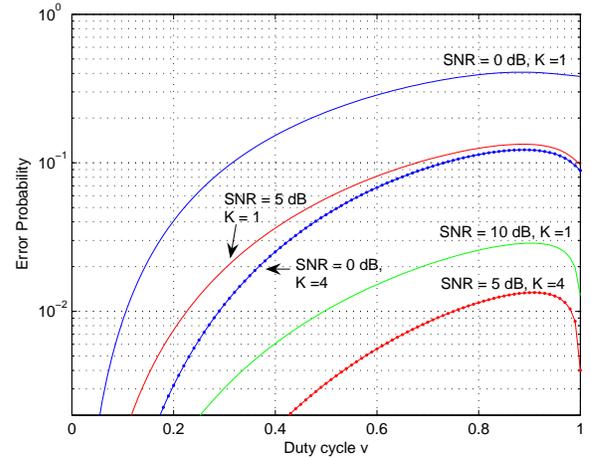}
\caption{Error probability vs. duty cycle $v$ for 8-OOFSK signaling over $L = 2$ independent Rician fading channels with equal Rician
factor. The curves for which Rician factor is $K = 1$ is plotted as solid lines while the curves for which $K =4$ are plotted as dotted-solid lines.  The receiver has only channel distribution knowledge. }\label{fig:Pe-vs-v}
\end{center}
\end{figure}

\begin{figure}
\begin{center}
\includegraphics[width = \figsize\textwidth]{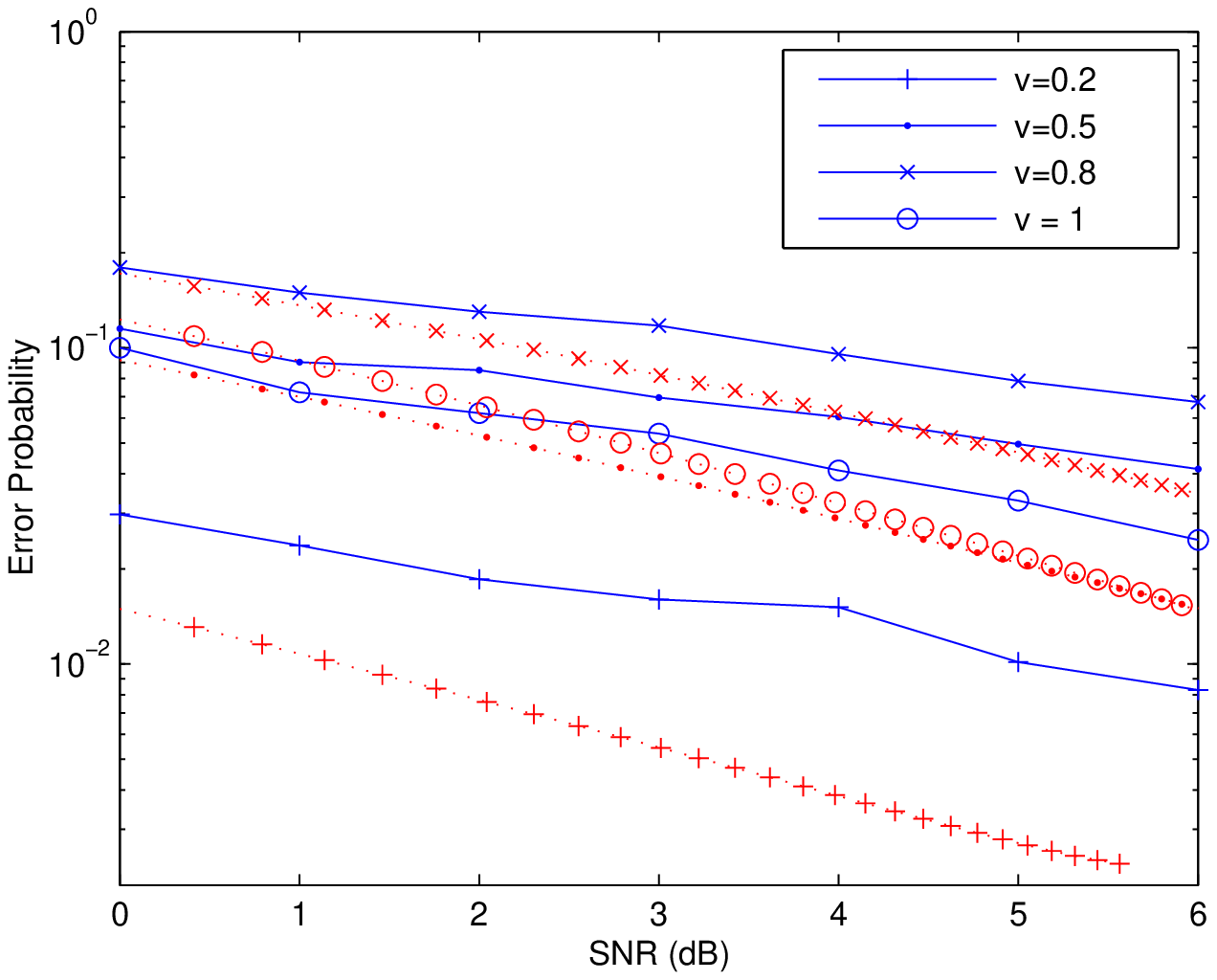}
\caption{Error probability vs. SNR for 4-OOFSK signaling over two
Rician fading channels (i.e., $ L = 2$)  with equal Rician
factor $K=\frac{1}{8}$. The receiver has only channel distribution knowledge. Curves with solid lines provide the error rates when the channels are correlated (correlation coefficient $\rho=\frac{1}{4}$) while dotted lines are for independent channels. Duty cycle values are 0.2, 0.5, 0.8, 1.} \label{3noncor}
\end{center}
\end{figure}
Next, we present the numerical results. We define the Rician factor
as $K= \frac{|E\{h_l\}|}{E\{|h_l - E\{h_l\}|^2\}}=
\frac{|d_l|^2}{\sigma^2}$ and correlation coefficient as
$\rho=\frac{\text{cov}(h_i,h_j)}{\sqrt{\text{var}(h_i)
\text{var}(h_j)}}$. Figure \ref{Pe-noncoh-M8}  plots the computed error probability curves for different duty cycle, $v$, values
when 8-OOFSK signals are transmitted over unknown independent Rician fading
channels all with the same Rician factor of $K = 1$. Solid lines provide the error rates for $L = 2$ channels while dashed lines are for $L = 8$ channels. Note that in
$M$-OOFSK modulation, the maximum number of bits that can be carried on the average
is equal to the entropy $ H(v) = v \log_2 (M/v)
+ (1 - v) \log_2 (1 / (1 - v))$ bits/symbol which decreases to zero as $v
\to 0$. Hence, decreasing the duty cycle diminishes the data rates. Therefore, for fair comparison, Fig. \ref{Pe-noncoh-M8} plots the curves as a function of the SNR normalized by the entropy of the $M$-OOFSK
signaling, giving the SNR per bit or equivalently $E_b/N_0$. In this figure, we observe that for sufficiently small $E_b/N_0$ values, the highest error rates are seen when $\nu = 1$ (i.e., when conventional FSK is employed). In fact, we can show that as $\text{SNR} \to 0$, the error probabilities approach $P_e \to v$ when $v < M/(M+1)$, and $P_e \to (M-v)/M$ when $v > M/(M+1)$ (see Appendix \ref{appendix}). Hence, asymptotically as SNR vanishes, error probabilities when $v < (M-1)/M$ are smaller than those attained when $v = 1$. Note that when $M = 8$, $(M-1)/M = 0.875$, confirming the above-mentioned observation in Fig. \ref{Pe-noncoh-M8}. However, we also see in this figure that as $E_b/N_0$ increases, conventional FSK (i.e., OOFSK with $v = 1$) achieves lower error rates than those achieved when $v = 0.8$ and $v = 0.5$. For instance, when $L =2$, conventional FSK out performs OOFSK with $v = 0.8$ and $v = 0.5$ when $E_b/N_0$ is greater than $-4.6$ dB and $4.4$ dB, respectively. Similar behavior is observed for the case of $L = 8$ when $E_b/N_0$ exceeds $-6.6$ dB  and $-1.3$ dB. Note that the impact of the minimum distance in the constellation on error rates is pronounced when SNR increases. Note further that for OOFSK modulation with $v < 1$, the minimum distance between the transmitted signals is proportional to $\sqrt{P/v}$ while the minimum distance for conventional orthogonal FSK is proportional to $\sqrt{2P}$. Hence, in contrast to the low-SNR regime, gains at moderate SNR levels are expected only if $v < 0.5$, as evidenced in the graphs of $v = 0.2$ and $\nu = 0.1$ where the duty factor is
sufficiently decreased, and hence consequently the minimum distance is increased. Note that when $\nu = 0.1$, we observe approximately an order of magnitude improvement with
respect to FSK modulation.
Fig. \ref{Pe-noncoh-M8} also demonstrates the diversity benefits of increasing the number of reception channel from $L = 2$ to $L = 8$. As SNR grows without bound, we can show that $P_e \to 0$ for all values of $v \in (0,1]$, and hence any possible gains in the error performance obtained by increasing the peakedness diminishes.

Fig. \ref{fig:Pe-vs-v} plots the
error probabilities as a function of the duty cycle parameter $v$ for 8-OOFSK signaling over two independent Rician channels with identical Rician factor values.  Fig. \ref{fig:Pe-vs-v} confirms the earlier observation that improvements in error rates at moderate-to-high SNR levels are realized if $v$ is sufficiently small. We observe that the values of $v$ required to produce improvements are in general dependent on SNR and the Rician factor $K$, and get smaller with increasing SNR and $K$. For instance, when SNR = 0 dB and $K = 1$, error rates lower than those achieved by FSK signaling is attained when $v < 0.77$. On the other hand, when SNR = 5 dB and $K = 4$, $v$ should be lowered below 0.53. Finally, Fig. \ref{3noncor} provides a comparison between the performances in correlated and independent channels. The error rates for correlated channels with correlation coefficient $\rho = 1/4$ are obtained through simulations (rather than computations) by using the decision rules derived for independent channels. In this figure, we immediately note the deterioration in the performance due to channel correlation.

\section{Noncoherent Detection of OOFSK with Instantaneous Channel Knowledge} \label{sec:coherent}

\subsection{Detection Rule}
In this section, we assume that the instantaneous values of $|h_l|$
for all $l$ are perfectly known to the receiver while the phases of
the fading coefficients, $\theta_h$, are still unknown. We further
assume that the transmitter has no knowledge of the fading
coefficients. Similarly as in the previous section, the receiver first computes
the energy of each frequency at each antenna through $R_{m,l} =
|Y_{m,l}|^2$. Conditioned on $|h_l|$ and the transmitted signal
$s_k$, the pdf of the chi-squared distributed $R_{l,m}$ is given by \cite{proakis}
\begin{align}
&f_{R_{l,m}|\,|h_l|,s_k}(R_{l,m})=\nonumber
\\
&\quad\left\{\begin{array}{ll}e^{-(R_{l,m}+A^2|h_l|^2)}I_0\left(2A|h_l|\sqrt{R_{l,m}}\right)&m=k\\
e^{-R_{l,m}}&m\neq k\end{array}\right..\label{eq:Rlmpdfcoh}
\end{align}
When we compare the statistics of the $R_{l,m}$ in Sections
\ref{sec:noncoherent} and \ref{sec:coherent} (i.e., compare
(\ref{eq:Rlmpdfnoncoh}) and (\ref{eq:Rlmpdfcoh})), we note that
(\ref{eq:Rlmpdfcoh}) can be obtained from (\ref{eq:Rlmpdfnoncoh}) by
assuming $\sigma^2 = 0$ and replacing $|d_l|$ by the random channel
magnitude $|h_l|$ in (\ref{eq:Rlmpdfnoncoh}). Due to this similarity
in the statistics, it can be seen that the results for the known
channel can be obtained as a special case of those in Section
\ref{sec:noncoherent} if we set $\sigma^2 = 0$ and replace $|d_l|$
by $|h_l|$ in the formulas in Section \ref{sec:noncoherent}. For
instance, we again assume here that the receiver combines, for each
frequency, the energies across the antennas, and obtain
$$R_m=\sum_{l=1}^LR_{l,m} = \sum_{l=1}^L|Y_{l,m}|^2.$$ It can be shown
that the conditional pdf of the vector of sum energies, $\R = [R_1,
\ldots, R_M]$, is given by \eqref{pdf_sm} on the next page
\begin{figure*}
\begin{align}\label{pdf_sm}
&f_{\R||\h|, s_k}(\R)=
\left\{\begin{array}{ll}\left(\frac{R_k}{\xi}\right)^{\frac{L-1}{2}}e^{-(R_k+\xi)}I_{L-1}(2\sqrt{R_k\xi})\prod_{n=1\
n\neq k
}^M \frac{R_n^{L-1}e^{-R_n}}{\Gamma(L)} &  k\neq 0\\
\frac{1}{\left[\Gamma(L)\right]^{M}}\prod_{n=1}^MR_n^{L-1}e^{-R_n}
& k=0\end{array}\right.
\end{align}
\hrule
\end{figure*}
where $\xi=A^2\sum_{l=1}^L|h_l|^2$, and $|\h| =
[|h_1|,\ldots,|h_L|]$ is the vector of the magnitudes of the fading
coefficients. Note that assuming $\sigma^2 = 0$ and replacing
$|d_l|$ by $|h_l|$ in (\ref{eq:pdf_R_noncoh}) also leads to
(\ref{pdf_sm}). Using this approach, one can easily verify that the
MAP rule that detects $s_k$ for $k \neq 0$ in known channels is
\begin{align}\label{finalcompare1}
\begin{split}
R_k&>R_m  \,\, \forall m \neq k\qquad \text{and}
\\
R_k&> \tau_2 =
\left\{
\begin{array}{ll}
g_2^{-1}(T_2) & \text{if } T_2 \ge \frac{\xi^{\frac{L-1}{2}}}{(L-1)!}
\\
0 & \text{otherwise}
\end{array}\right.
\end{split}
\end{align}
where $g_2(R_k) = R_k^{-\frac{L-1}{2}}I_{L-1}(2\sqrt{R_k\xi})$ and $T_2=\frac{M(1-v)e^\xi \xi^{\frac{L-1}{2}}}{v(L-1)!}$. Note
that (\ref{finalcompare1}) is the rule that detects the signal
$s_k(t)$ for $k \neq 0$. The zero signal $s_0(t)$ is detected if $
R_k < \tau$ $\forall k. $

\subsection{Probability of Error}
We first assume that $s_1(t)$ is the transmitted signal.
Then, specializing the results in Section \ref{subsec:errornoncoh}, we find the correct detection probability as
\begin{align}
P_{c,1}&=P(R_1>R_2,R_1>R_3,\ldots,R_1>R_M,R_1>\tau_2|s_1,|\h|) \nonumber \\
&=\int_{\tau_2}^\infty\left[1-e^{-x}\sum_{l=0}^{L-1}\frac{x^l}{l!}\right]^{M-1}f_{R_1|
|\h|, s_1}(x)dx
\\
&=\sum_{n=0}^{M-1}(-1)^n\left(\!\!\!\begin{array}{cc}M-1\\n\end{array}\!\!\!\right)\nonumber
\\
&\quad\sum_{i=0}^{n(L-1)}c_{in}\int_{\tau_2}^\infty
x^i\left(\frac{x}{\xi}\right)^{\frac{L-1}{2}}
e^{-[(n+1)x+\xi]}I_{L-1}(2\sqrt{x\xi})dx.
\end{align}
Note that if we specialize (\ref{eq:errorprobhyper}), an expression involving a hypergeometric function can also be obtained. The probability of correct detection when signal $s_0(t)$ is
transmitted is
\begin{align}
P_{c,0}&=P\left(R_1<\tau_2,\ldots,R_M<\tau_2||\h|,s_0\right)
\\
&   =
\left( 1 - e^{-\tau_2} \sum_{l=0}^{L-1}
\frac{\tau_2^l}{l!}\right)^M.
\end{align}
Hence, the probability of error as a function of the instantaneous
signal-to-noise ratio is
\begin{align} \label{eq:proboferror}
P_e=1-(vP_{c,1}+(1-v)P_{c,0}).
\end{align}
Since the channel is assumed to be known, error probability in
(\ref{eq:proboferror}) is a function of the fading coefficients
through $\chi=\sum_{l=1}^{L}|h_l|^2$. Hence, the average probability
of error is obtained by computing $ \bar{P}_e=\int_0^\infty P_e \,
f_{\chi}(\chi)d\chi. $ If $h_l$ is a complex Gaussian random
variable with mean value $d_l$ and variance $\sigma^2$ and $\{h_l\}$
are mutually independent, $\chi$ is a chi-square random variable
with $2L$ degrees of freedom and has a pdf given by $
f_{\chi}(\chi)=\frac{1}{\sigma^2}\left(\frac{\chi}{s^2}\right)^{\frac{L-1}{2}}e^{-\frac{\chi+s^2}{\sigma^2}}
I_{L-1}\left(\frac{2\sqrt{\chi s^2}}{\sigma^2}\right) $ where,
$s^2=\sum_{l=1}^{L}|d_l|^2$.

When the fading coefficients $\{h_l\}$ are correlated, the average
error probability $\bar{P}_e$ can be obtained by evaluating the
expected value of $P_e$ with respect to the joint distribution of
$(|h_1|, \ldots, |h_L|)$, which involves an $L$-fold integration.
However, if $\{|h_l|\}$ are Nakagami-$m$ distributed, closed-form
expressions for $f_\chi(\chi)$ are provided in \cite{george1}, which
lead to a single integration.

\begin{figure}
\begin{center}
\includegraphics[width = \figsize\textwidth]{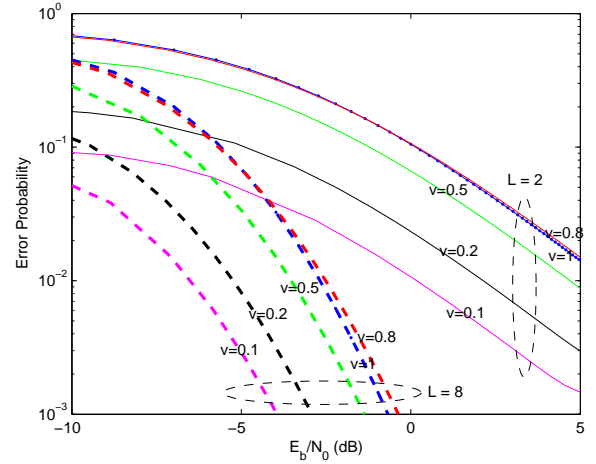}
\caption{Error probability vs. $E_b/N_0$ (dB) for 8-OOFSK signaling over independent Rician fading channels with equal Rician
factor $K=1$. The receiver has the knowledge of the instantaneous values of the fading magnitude. Solid lines provide the error rates for $L = 2$ channels while dashed lines are for $L = 8$ channels. Duty cycle values are 0.1, 0.2, 0.5, 0.8, 1.}\label{2uncoco}
\end{center}
\end{figure}

\begin{figure}
\begin{center}
\includegraphics[width = \figsize\textwidth]{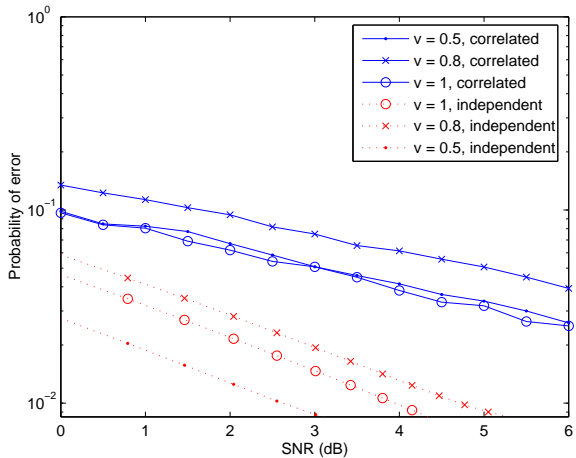}
\caption{Error probability vs. SNR for 4-OOFSK signaling over two
correlated known Rician fading channels with equal Rician factor
$K=\frac{1}{8}$ and correlation coefficient $\rho=\frac{1}{4}$. Duty
factor values are $v = 1, 0.8, 0.5$, and $0.2$.}\label{2coco}
\end{center}
\end{figure}

\begin{figure}
\begin{center}
\includegraphics[width = \figsize\textwidth]{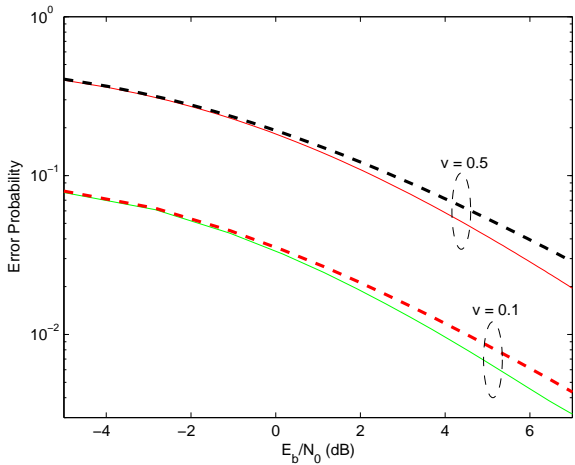}
\caption{Error probability vs. $E_b/N_0$ (dB) for 8-OOFSK signaling over two independent Rayleigh fading channels, i.e., $L = 2$. Solid curves provide the error rates when the fading magnitude is perfectly known at the receiver while dashed curves are the error rates when the receiver has only channel distribution knowledge.}\label{coh-vs-noncoh}
\end{center}
\end{figure}
Figure \ref{2uncoco} plots, for different values of $v$ and different number of reception channels, the computed error probability vs. $E_b/N_0$ curves for 8-OOFSK over known independent Rician fading channels each with $K = 1$. Conclusions similar to that in Section \ref{sec:noncoherent} can be immediately drawn here. However, note that error performance improves even when $v = 0.5$ in this case.  When $v = 0.2$ and 0.1, even lower error probabilities are attained. Equivalently, we can conclude that for fixed error rates, substantial energy gains are realized, rendering OOFSK signaling a very
energy efficient transmission technique.  In Fig. \ref{2coco}, a comparison between the performances in correlated and independent channels is given. Although the performance is
deteriorated due to correlation, OOFSK with sufficiently small duty
factor still considerably improves the error performance. Finally, in Fig. \ref{coh-vs-noncoh}, we plot the error probabilities for 8-OOFSK signaling with $v = 0.5$ and $v = 0.1$ over both known and unknown Rayleigh fading channels. We note that knowing the fading magnitudes does not provide benefits at low $E_b/N_0$ values. At relatively high values of the bit energy, approximately 1 dB gain in $E_b/N_0$ can be achieved for fixed error rates. Note that in known and unknown channels, detection rules are similar and hence the implementation complexities of the detectors are comparable. However, in order to get the performance results of the known channels, channels have to be estimated at the receiver. Therefore, gains in the performance should be compared with the resources expended for channel estimation to identify the net gains.

\section{Conclusion} \label{sec:conclusion}

In this paper, we have analyzed the error performance of OOFSK
modulation in both known and unknown fading channels when the receiver is equipped with
multiple antennas. The receiver is assumed to obtain the sum energy
for each frequency by combining the energies of that frequency
present at each antenna. We have identified the MAP detection rules
as comparisons of the sum energies of frequencies with each other
and a certain threshold, and obtained analytical error probability
expressions. Through numerical and simulation results, we have shown the benefits
of increasing the peakedness of the signals, and quantified the
impacts of the number of antennas, antenna correlation, duty cycle values, and unknown
channel conditions on the error performance.

\appendices

\section{} \label{appendix}

In obtaining the asymptotic values of the error probabilities, the critical step is to find the value of the threshold $\tau_1$ in \eqref{ncfinalcompare1} as SNR vanishes. When $v < \frac{M}{M+1}$, we can see from the definitions in \eqref{ncfinalcompare1} that
\begin{gather} \label{eq:lowsnrcond}
\lim_{\text{SNR} \to 0} \frac{g_1(\tau_1)}{T_1} = 1
\end{gather}
from which we can easily deduce that $\tau_1 \to \infty$ as SNR decreases. Then, from the correct detection probabilities in \eqref{eq:Pc1noncoh} and \eqref{eq:Pc0noncoh}, we have
$\lim_{\text{SNR} \to 0} P_{c,1} = 0$ and $\lim_{\text{SNR} \to 0} P_{c,0} = 1$, which lead to $$\lim_{\text{SNR} \to 0} P_e = \lim_{\text{SNR} \to 0} (1-v P_{c,1} - (1-v)P_{c,0}) = v.$$

On the other hand, when $v > \frac{M}{M+1}$, we can find again from \eqref{ncfinalcompare1} that $\tau_1 \to 0$ as SNR $\to 0$. In this case, we readily obtain that
$
\lim_{\text{SNR} \to 0} P_{c,1} = \frac{1}{M}$ and $
\lim_{\text{SNR} \to 0} P_{c,0} = 0$,
which lead to
$$
\lim_{\text{SNR} \to 0} P_e = \lim_{\text{SNR} \to 0} (1-v P_{c,1} - (1-v)P_{c,0}) = 1 - \frac{v}{M} = \frac{M-v}{M}.
$$

\end{document}